\documentclass[12pt]{article}

\begin{document}
\begin{center}
{\bf On Torsion Fields in Higher Derivative Quantum Gravity}

\vspace{5mm}
 S.I. Kruglov\\

\vspace{5mm}
\textit{University of Toronto at Scarborough,\\ Physical and Environmental Sciences Department, \\
1265 Military Trail, Toronto, Ontario, Canada M1C 1A4}
\end{center}

\begin{abstract}
We consider axial torsion fields which appear in higher derivative
quantum gravity. It is shown, in general, that the torsion field
possesses states with two spins, one and zero, with different
masses. The first-order formulation of torsion fields is performed.
Projection operators extracting pure spin and mass states are given.
We obtain the Lagrangian in the framework of the first order
formalism and energy-momentum tensor. The effective interaction of
torsion fields with electromagnetic fields is discussed. The
Hamiltonian form of the first order torsion field equation is given.
\end{abstract}

\section{Introduction}
It is obvious nowadays that the Standard Model (SM) should be
extended by including gravity. However, classical General Relativity
(GR) based on the Einstein-Hilbert action
\begin{equation}
S=-\frac{1}{\kappa^2}\int d^4x\sqrt{-g}R,
  \label{1}
\end{equation}
where $R$ is a scalar curvature of space-time, is a nonrenormazible
theory. The requirement of renormalizability  of the theory of
gravity leads to the Lagrangian with quadratic terms in curvature
($R^2$ theory of gravity) \cite{Stelle}. Although the quantum theory
of gravity becomes a renormazible theory, the new difficulty appears
- nonunitarity. This fact is connected with higher derivative
equations of motion. Field equations in $R^2$ theory of gravity are
forth-derivative equations. Now much attention is paid to higher
derivative quantum field theories. For example, the higher
derivative scalar field theory was considered in \cite{Kruglov}, and
higher derivative fermionic field equations were discussed in
\cite{Dvoeglazov}, \cite{Kruglov1}. The nonunitarity is the main
obstacle to the consistent theory of quantum gravity. Possibly this
difficulty will be solved by developing more a fundamental unified
theory (the theory of everything, M-theory, the string theory, etc.)
Originally, GR was applied for macroscopic objects, and the
energy-momentum tensor of matter is the source of gravitational
field characterized by the metric tensor $g_{\mu\nu}$ (the Riemann
space-time). But it is possible to construct the microscopic theory
of gravity where the spin of elementary particles is a source of
another characteristic of space-time $-$torsion. Torsion fields are
additional degrees of freedom that are independent of metric. This
natural extension of gravity is gravity with torsion that is
described by the Riemann-Cartan space-time $ \textit{U}_4$
\cite{Hehl}. The gravity of macroscopic objects can then be
considered by averaging the microscopic equations over all volume.
Therefore, we can naturally include torsion fields in the quantum
theory of gravity. Torsion fields interact with the spin of
particles and may contribute to physical observations.

It should be mentioned that the quantum gravity plays a very
important role in the investigation of early universe cosmology. In
cosmology, there are some puzzles such as dark energy, positive
cosmological constant etc. Possibly, the torsion gravity may solve
some problems in cosmology, and therefore, the generalization of
metric gravity by including torsion is justified.

In this paper, we pay attention to torsion fields which appear in
higher derivative quantum gravity. As the gravitation interaction
(connected with the metric tensor $g_{\mu\nu}$) between elementary
particles is extremely weak, one can consider the flat Minkowski
space-time with torsion.

 The paper is organized as follows. In section 2 we outline the geometry
 of space-time with torsion and write down the action of higher derivative
quantum gravity with torsion. The free torsion field in flat
space-time is considered in Sec. 3. We formulate the first order
wave equation for the torsion neutral field in the 11-dimensional
matrix form. It is shown that the torsion field which appears in
higher derivative quantum gravity possesses two spins, one and zero,
with different masses. When the mass of the spin-zero state
approaches to infinity, one comes to the Proca equation. Projection
operators extracting pure spin and mass states are given. We obtain
the Lagrangian in the framework of the first order formalism and the
energy-momentum tensor. The possible effective interaction of the
torsion field with external electromagnetic fields is discussed in
Sec.4. The Hamiltonian form of an equation is given. We make a
conclusion in Sec.5.

\section{Higher derivative quantum gravity}

\subsection{Connection and torsion}

Space-time represents the four-dimensional manifold $X_4$ with
points $x^\mu$, $\mu=0,1,2,3$, and $x^0$ is the time. The parallel
transfer of a contravariant vector $V^\mu$ is given by \cite{Hehl}
\begin{equation}
dV^\mu=-\Gamma^\mu_{\alpha\beta}V^\alpha dx^\beta, \label{2}
\end{equation}
where $\Gamma^\mu_{\alpha\beta}$ is the affine connection with 64
components. The affine connection can be written as a sum of the
symmetric part $\Gamma^\mu_{(\alpha\beta)}$ and the antisymmetric
part $\Gamma^\mu_{[\alpha\beta]}$:
\begin{equation}
\Gamma^\mu_{\alpha\beta}=\Gamma^\mu_{(\alpha\beta)}+\Gamma^\mu_{[\alpha\beta]},
~~\Gamma^\mu_{[\alpha\beta]}=\frac{1}{2}\left(\Gamma^\mu_{\alpha\beta}
-\Gamma^\mu_{\beta\alpha}, \right)\label{3}
\end{equation}
so that the antisymmetric part transforms as a tensor but the
symmetric part not. The antisymmetric part of the connection
$S^{..\mu}_{\alpha\beta}\equiv \Gamma^\mu_{[\alpha\beta]}$  is
called the Cartan torsion tensor and it possesses 24 components. In
GR only the $\Gamma^\mu_{(\alpha\beta)}$ is explored with the
additional requirements $\Gamma^\mu_{[\alpha\beta]}=0$,
$\nabla_\alpha g_{\mu\nu}=0$, and the covariant derivative
$\nabla_\alpha$ is defined as
\begin{equation}
\nabla_\alpha V^\beta =\partial_\alpha V^\beta
+\Gamma^\beta_{\alpha\mu}V^\mu . \label{4}
\end{equation}
The symmetric part of the connection, in GR, is named the
Christoffel symbol and is expressed via the metric tensor:
\begin{equation}
\left\{\begin{array}{c}
\mu\\
\alpha\beta
\end{array}\right\}
=\frac{1}{2}g^{\mu\lambda}\left(\partial_\alpha
g_{\lambda\beta}+\partial_\beta g_{\lambda\alpha}-\partial_\lambda
g_{\alpha\beta}\right). \label{5}
\end{equation}
The square of the infinitesimal interval is given by
$ds^2=-g_{\mu\nu}dx^\mu dx^\nu$. In the Minkowskian flat space-time
$g_{\mu\nu}=\eta_{\mu\nu}=$diag$(-1,+1,+1,+1)$.

Now, we imply that the affine connection $\Gamma^\mu_{\alpha\beta}$
is not symmetric (connection with torsion), but the metric tensor
$g_{\mu\nu}$ is still constrained by the equation
$\widetilde{\nabla}_\alpha g_{\mu\nu}=0$ (nonmetricity =0), where
$\widetilde{\nabla}$ being the covariant derivative with torsion.
The contorsion tensor $K_{\alpha\beta}^{..\mu}$ is introduced by the
relation \cite{Hehl}:
\begin{equation}
\Gamma^\mu_{\alpha\beta}=\left\{\begin{array}{c} \mu\\\alpha\beta
\end{array}\right\}-K_{\alpha\beta}^{..\mu},~~K_{\alpha\beta}^{..\mu}
=-S^{..\mu}_{\alpha\beta}+ S^{.\mu ,}_{\beta .\alpha}-
S^{\mu}_{.\alpha\beta}= -K^{.\mu}_{\alpha .\beta} .\label{6}
\end{equation}
As usual, indices are raised and lowered with the help of the metric
tensor. It should be mentioned that the contorsion tensor
$K_{\alpha\beta}^{..\mu}$ possesses 24 components and depends on
metric and torsion tensors \cite{Hehl}. The reader can find the
classification of the torsion tensor in \cite{Hehl1},
\cite{Capozziello}.

\subsection{Action of renormalized quantum gravity with torsion}

The action of renormalized quantum gravity with torsion is given by
(see \cite{Buchbinder} and references therein)
\[
S=\int d^4x\sqrt{-g}\biggl(\frac{1}{2\lambda}C^2_{\mu\nu\alpha\beta}
-\frac{\omega}{3\lambda}R^2-\frac{1}{\kappa^2}R+\Lambda
+\alpha_1S_{\mu\nu}^2
\]
\begin{equation}
+\alpha_2(\nabla_\mu S^\mu)^2+\alpha_3(S_\mu S^\mu)^2
+\alpha_4R^{\mu\nu}S_\mu S_\nu,
  \label{7}
\end{equation}
\[
+\alpha_5RS_\mu S^\mu-\frac{1}{\kappa^2}\xi S_\mu S^\mu\biggr),
\]
where $C_{\mu\nu\alpha\beta}$ is the Weyl tensor that is expressed
through the curvature tensor, $R_{\mu\nu}$ is the Ricci tensor,
$\Lambda$ is the cosmological constant, $S_{\mu\nu}=\partial_\mu
S_\nu-\partial_\nu S_\mu $, and the axial vector $S_\mu$ is the
antisymmetric part of the torsion tensor $S^{..\mu}_{\alpha\beta}$:
$S^\mu=(1/3!)\epsilon^{\mu\nu\alpha\beta}S_{\nu\alpha\beta}$.
Constants $\kappa$, $\omega$, $\lambda$, $\xi$, $\alpha_i$
characterize the gravitational interaction.

We note that in Eq.(7) all terms appear as counterterms within the
renormalization procedure. Therefore, one can not ignore some terms
in Eq.(7) to have the renormalized quantum theory. But there are
still two difficulties in the quantum theory based on the action
(7). One problem is connected with the cosmological constant. To
quantize the metric, we need to expand the flat background, but the
flat metric $\eta_{\mu\nu}$ is not the solution of equations of
motion. If one puts $\Lambda=0$ in the bar action, a divergent
counterterm appears from the loop calculations. The second problem
is the unitarity problem. This is the main difficulty in the higher
derivatives quantum theory. In such theories the ghosts appear with
the negative norm that are considered as non-physical states.
Nonunitarity of $R^2$-gravity shows that the gravity theory with the
action (7) may be considered as an approximation (or the effective
model) of the fundamental theory. The interaction of torsion with
matter fields is discussed in \cite{Buchbinder}, \cite{Shapiro}. It
should be mentioned that the axial vector $S_\mu$ interacts with all
fermions, but other components of torsion fields $S_{\mu\nu\alpha}$
can interact with matter only if one introduces non-minimal
couplings \cite{Shapiro}. This interaction can not appear at the
quantum level if we consider only the minimal interaction with
torsion fields.

The generalization of the Einstein -Hilbert action (1) on the case
of gravity with torsion leads to the Einstein-Cartan theory. In this
case the torsion fields do not propagate because there are no
kinetic terms in the Lagrangian. The effect of torsion fields is in
contact spin-spin interactions. At low energies this effect is
negligible and can be important for physics in the Early Universe.
For example, the problem of singularity can be solved in the
Einstein-Cartan theory. The main shortcoming of the Einstein-Cartan
theory is its non-renormalizability.

In the renormalized quantum gravity kinetic terms of torsion fields
appear due to loop calculations, and therefore they have to present
in the bare Lagrangian. To investigate the mass spectrum and
properties of propagating torsion fields, we consider here, for
simplicity, the flat space-time ($g_{\mu\nu}=\eta_{\mu\nu}$) with
torsion. So, we neglect the effect of the metric. For this purpose
only quadratic terms in the torsion field $S_\mu$ in Eq.(7) are
important. From Eq.(7), we arrive at the action for the free torsion
field
\begin{equation}
S_T=\int d^4x\biggl(\alpha_1S_{\mu\nu}^2+\alpha_2(\partial_\mu
S^\mu)^2 -\frac{1}{\kappa^2}\xi S_\mu S^\mu\biggr).
  \label{8}
\end{equation}
We have ignored the self-interaction term $\alpha_3(S_\mu S^\mu)^2$
which is in fourth order in fields. The particular case $\alpha_2=0$
was considered in \cite{Shapiro}, \cite{Shapiro1}, and the case
$\alpha_1=0$ was discussed in \cite{Carroll}. It should be mentioned
that all terms in (8) have to be presented in order to preserve
renormalizability of quantum gravity with torsion. The requirement
$\alpha_2=0$ to have unitarity in the torsion sector is a constraint
because even without a torsion the quantum gravity is non-unitary
theory. Therefore, we investigate here the general case of non-zero
coefficients in Eq.(8).

\section{ Propagating torsion fields}

\subsection{First order formulation}

By renormalization of the field $S_\mu$, we can make the standard
kinetic term in action (8): $(-1/4)S_{\mu\nu}^2$. This means that
without loss of generality, one may put $\alpha_1=-1/4$. The
corresponding Lagrangian was already considered in \cite{Kruglov2}
for complex fields describing charged particles. The torsion fields
$S_\mu$ are real, and correspond to neutral fields, but it does not
matter for the analysis of the mass spectrum. It was shown that
fields described by the action (8) possess the state with spin-1 and
the square mass $m^2=2\xi/\kappa^2$ and the state with spin-0 and
the square mass $m_0^2=-m^2/2\alpha_2$. Then the Lagrangian
corresponding to the action (8) becomes
\begin{equation}
\mathcal{L}_T=-\frac{1}{4}S_{\mu\nu}^2-\frac{m^2}{2m_0^2}(\partial_\mu
S^\mu)^2 -\frac{1}{2}m^2 S_\mu S^\mu.
  \label{9}
\end{equation}
From the Lagrangian (9), one can obtain field equations as follows:
\begin{equation}
\partial^\mu\partial_\mu S_\nu +\left( \frac{m^2}{m_0^2}-1\right) \partial _\nu \left(
\partial^\alpha S_\alpha \right) -m^2 S_\nu =0.  \label{10}
\end{equation}
At the case $m_0\rightarrow \infty$, one arrives at the Proca
equation describing pure spin-1 states. If $m=m_0$, we come to the
case investigated in \cite{Kruglov3}.

It is convenient formally to go to Euclidian space-time
($\eta_{\mu\nu}\rightarrow \delta_{\mu\nu}$) by introducing fourth
components of vectors $S_4=iS_0$. At this convention there is no
difference between contra - and covariant components of vectors, and
$\partial _\mu =(\partial_m,\partial_4 $) (m=1,2,3),
$\partial_4=\partial/\partial(it)$.

Eq.(10) may be represented in the first order formalism as
\cite{Kruglov2}
\begin{equation}
\left( \alpha _\mu \partial _\mu +m P_1+\frac{m^2_0}{m} P_0\right)
\Psi (x)=0, \label{11}
\end{equation}
where
\begin{equation}
\Psi (x)=\left\{ \psi _A(x)\right\} =\left(
\begin{array}{c}
(1/m)S(x) \\
S _\mu (x) \\
(1/m)S _{\mu \nu}(x)
\end{array}
\right) \hspace{0.5in}(A=0, \mu , [\mu \nu ]), \label{12}
\end{equation}
$S (x)=-(m^2/m_0^2)\partial_\mu S_\mu(x)$, and matrices are given by
\[
\alpha _\mu =\beta _\mu ^{(1)}+\beta _\mu ^{(0)},~~\beta _\nu
^{(1)}=\varepsilon ^{\mu ,[\mu \nu ]}+\varepsilon ^{[\mu \nu ],\mu
},~~\beta _\nu ^{(0)}=\varepsilon ^{\nu ,0}+\varepsilon ^{0,\nu },
\]
\vspace{-8mm}
\begin{equation}
\label{13}
\end{equation}
\vspace{-8mm}
\[
P_1=\varepsilon ^{\mu ,\mu }+\frac 12\varepsilon ^{[\mu \nu ],[\mu
\nu ]},\hspace{0.5in}P_0=\varepsilon ^{0,0},
\]
expressed trough the elements of the entire matrix algebra
$\varepsilon ^{A,B}$ with the properties: $\left( \varepsilon
^{A,B}\right) _{CD}=\delta _{AC} \delta _{BD}$, $\varepsilon
^{A,B}\varepsilon ^{C,D}= \delta _{BC}\varepsilon^{A,D}$. Matrices
$\beta _\mu ^{(1)}$ and $\beta _\mu ^{(0)}$ are defined in $10-$ and
$5-$dimensional subspaces and obey the Petiau-Duffin-Kemmer algebra.
The projection matrices $P_1$,$P_0$ satisfy the relations:
\begin{equation}
P_1^2=P_1,\hspace{0.5in}P_0^2=P_0,\hspace{0.5in}P_1 P_0=0,
\hspace{0.5in}P_1+P_0=1. \label{14}
\end{equation}

The relativistic first order $11\times 11$-matrix eqution (11)
describes torsion fields possessing two spins $0,$ $1$ with masses,
$m_0$ and $m$, respectively. The $11-$dimensional matrices $\alpha
_\mu $ obey the algebra as follows:
\[
\alpha _\mu \alpha _\nu \alpha _\alpha +\alpha _\alpha \alpha _\nu
\alpha _\mu +\alpha _\mu \alpha _\alpha \alpha _\nu +\alpha _\nu
\alpha _\alpha \alpha _\mu +\alpha _\nu \alpha _\mu \alpha _\alpha
+\alpha _\alpha \alpha _\mu \alpha _\nu =
\]
\begin{equation}
=2\left( \delta _{\mu \nu }\alpha _\alpha +\delta _{\alpha \nu
}\alpha _\mu +\delta _{\mu \alpha }\alpha _\nu \right). \label{15}
\end{equation}

The projection operators\footnote{In \cite{Kruglov2} more
complicated operator extracting states with spin-0 was constructed.}
\begin{equation}
M^{(1)}_{\varepsilon} =\frac{i\widehat{p}^{(1)}\left(
i\widehat{p}^{(1)}-\varepsilon m\right) }{ 2m^2},~~
M^{(0)}_{\varepsilon} =\frac{i\widehat{p}^{(0)}\left(
i\widehat{p}^{(0)}-\varepsilon m_0\right) }{ 2m_0^2},  \label{16}
\end{equation}
where $\widehat{p}^{(1)}=\beta^{(1)} _\mu p_\mu $,
$\widehat{p}^{(0)}=\beta^{(0)}_\mu p_\mu $, $p_\mu =({\bf p},ip_0)$
is the momentum of a particle, extract states with the mass $m$ and
$m_0$. Values $\varepsilon =1$, $\varepsilon =-1$ correspond to the
positive and negative energies of a particle, respectively. The
projection matrices $M^{(1,0)}_\varepsilon$ obey the relationships
\begin{equation}
M_\varepsilon^2=M_\varepsilon,
~~M^{(1)}_{\varepsilon}M^{(0)}_{\varepsilon}=
M^{(0)}_{\varepsilon}M^{(1)}_{\varepsilon}=0. \label{17}
\end{equation}

\subsection{Spin operators}

With the help of generators of the Lorentz group in the
$11-$dimensional representation space of the wave function
\begin{equation}
J_{\mu \nu }=\beta _\mu ^{(1)}\beta _\nu ^{(1)}-\beta _\nu
^{(1)}\beta _\mu ^{(1)},  \label{18}
\end{equation}
we obtain the squared spin operator (the factor 1/2 was missed in
\cite{Kruglov2})
\begin{equation}
\sigma ^2=\left( \frac 1{2m}\varepsilon _{\mu \nu \alpha \beta
}p_\nu J_{\alpha \beta }\right) ^2=\frac 1{m^2}\left(
\frac{1}{2}J_{\mu \nu }^2p^2-J_{\mu \sigma }J_{\nu \sigma }p_\mu
p_\nu \right), \label{19}
\end{equation}
obeying the ``minimal" equation
\begin{equation}
\sigma ^2\left( \sigma ^2-2\right) =0.  \label{20}
\end{equation}
The projection operators extracting states with spin-1 and spin-0
are given by \cite{Kruglov2}
\begin{equation}
S_{(0)}^2=1-\frac{\sigma ^2}2,\hspace{0.5in}S_{(1)}^2=\frac{\sigma
^2}2 \label{21}
\end{equation}
which obey equations $S_{(0)}^2S_{(1)}^2=0$, $\left(
S_{(0)}^2\right) ^2=S_{(0)}^2$, $\left( S_{(1)}^2\right)
^2=S_{(1)}^2$, $S_{(0)}^2+S_{(1)}^2=1$. The operators of the spin
projection on the direction of the momentum $\mathbf{p}$ are
\begin{equation}
\sigma _p=-\frac i{2\mid \mathbf{p}\mid }\epsilon
_{abc}p_aJ_{bc}=-\frac i{\mid \mathbf{p}\mid }\epsilon
_{abc}p_a\beta _b^{(1)}\beta _c^{(1)}, \label{22}
\end{equation}
and satisfy the matrix equation:
\begin{equation}
\sigma _p\left( \sigma _p-1\right) \left( \sigma _p+1\right) =0.
\label{23}
\end{equation}
Using the standard procedure, we find the projection operators
corresponding to spin projection one and zero \cite{Kruglov2}:
\begin{equation}
\widehat{S}_{(\pm 1)}=\frac 12\sigma _p\left( \sigma _p\pm 1\right),
\hspace{0.5in}\widehat{S}_{(0)}=1-\sigma _p^2. \label{24}
\end{equation}
The projection matrices extracting states with pure spin, spin
projection and energy are
\[
\Delta _{\varepsilon ,\pm 1}=M^{(1)}_\varepsilon
S_{(1)}^2\widehat{S}_{(\pm 1)}= \frac{i\widehat{p}^{(1)}\left(
i\widehat{p}^{(1)}-\varepsilon m\right) }{2m^2}\frac 12\sigma
_p\left( \sigma _p\pm 1\right),
\]
\begin{equation}
\Delta _\varepsilon ^{(1)}=M^{(1)}_\varepsilon
S_{(1)}^2\widehat{S}_{(0)}=\frac{i \widehat{p}^{(1)}\left(
i\widehat{p}^{(1)}-\varepsilon m\right) }{2m^2}\frac{\sigma ^2}
2\left( 1-\sigma _p^2\right), \label{25}
\end{equation}
\[
\Delta _\varepsilon ^{(0)}=M^{(0)}_\varepsilon
S_{(0)}^2\widehat{S}_{(0)} =\frac{i\widehat{p}^{(0)}\left(
i\widehat{p}^{(0)}-\varepsilon m_0\right) }{2m_0^2} \left( 1-\frac{
\sigma ^2}2\right) \left( 1-\sigma _p^2\right).
\]
Operators (25) also represent the density matrices for pure spin
spates. More information about spin operators the reader may find in
\cite{Kruglov2}.

\subsection{Energy-momentum tensor}

The Hermitianizing matrix is given by \cite{Kruglov2}
\begin{equation}
\eta =-\varepsilon ^{0,0}+\varepsilon ^{m,m}-\varepsilon
^{4,4}+\varepsilon ^{[m4],[m4]}-\frac{1}{2}\varepsilon ^{[m n ],[m
n]}, \label{26}
\end{equation}
and obey equations: $\eta \alpha _i=-\alpha^+ _i\eta^+$, $\eta
\alpha _4=\alpha^+ _4\eta^+$ ($i=1,2,3$). This matrix allows us to
obtain the relativistically invariant bilinear form $\overline{\Psi
}\Psi =\Psi ^{+}\eta \Psi$. Applying the Hermitian conjugation to
Eq.(11), and using equalities $[P_1,\eta]=[P_0,\eta]=0$, one obtains
the equation as follows:
\begin{equation}
\overline{\Psi} (x)\left( \alpha _\mu \overleftarrow{\partial} _\mu
-m P_1-\frac{m^2_0}{m} P_0\right) =0. \label{27}
\end{equation}

Now the Lagrangian, in the framework of the first order formalism,
reads
\begin{equation}
\mathcal{L}=-\overline{\Psi}(x)\left( \alpha _\mu \partial _\mu +m
P_1+\frac{m^2_0}{m} P_0\right) \Psi (x).
  \label{28}
\end{equation}
It should be noted that the functions $\overline{\Psi}(x)$ and $\Psi
(x)$ are dependent because the fields $(\textbf{S},S_0)$ are real.
Therefore, one has to make the variation of the action corresponding
to the Lagrangian (28) on $\overline{\Psi}(x)$ or $\Psi (x)$. With
the aid of the general expression for the canonical energy-momentum
tensor
\begin{equation}
T^c_{\mu\nu}=\frac{\partial \mathcal{L}}{\partial(\partial_\mu \Psi
(x))}\partial_\nu \Psi (x)-\delta_{\mu\nu}\mathcal{L}, \label{29}
\end{equation}
one obtains
\begin{equation}
T^c_{\mu\nu}= -\overline{\Psi} (x)\alpha_\mu
\partial_\nu\Psi (x). \label{30}
\end{equation}
We took into consideration that $\mathcal{L}=0$ for fields obeying
equations of motion (11). With the help of equations (11), (27), it
easy to verify that the canonical energy-momentum tensor is
conserved: $\partial_\mu T^c_{\mu\nu}=0$. Using Eq.(12), (26), one
finds $\overline{\Psi} (x)=\left(-(1/m)S(x),S_\mu
(x),-(1/m)S_{\mu\nu}(x)\right)$, and taking into account Eq.(13), we
obtain
\begin{equation}
T^c_{\mu\nu}=\frac{1}{m}\left(S\partial_\nu S _\mu-S_\mu\partial_\nu
S +S_\rho\partial_\nu S_{\mu\rho} -S_{\mu\rho}\partial_\nu
S_\rho\right). \label{31}
\end{equation}
The canonical energy-momentum tensor is not symmetric but can be
symmetrized by the standard procedure. For example, the symmetric
energy-momentum tensor may be obtained by varying the action on the
metric tensor. See also \cite{Dvoeglazov1} for the comparison. We
note that the identity $j_\mu=i\overline{\Psi} (x)\alpha_\mu \Psi
(x)=0$ is valid because the torsion fields $(\textbf{S},S_0)$
describe neutral fields and the electric current is equal to zero,
$j_\mu=0$.

\section{Electromagnetic interactions of torsion fields}

\subsection{Non-minimal interactions}

Torsion fields can interact with electromagnetic fields only
non-minimally because they do not carry the electric charge. In
\cite{Buchbinder} the interaction of the kind $a(\partial_\mu S_\nu)
\widetilde{\mathcal{F}}^{\mu\nu}$ in the Lagrangian was suggested,
where $\widetilde{\mathcal{F}}^{\mu\nu}$ is the dual tensor of
electromagnetic fields. However, this interaction results in the
appearance of the additional term
$a\partial_\mu\widetilde{\mathcal{F}}_{\mu\nu}$ in Eq.(10) which is
non-zero only in the presence of magnetic monopoles.

Here, we consider non-trivial electromagnetic interaction of torsion
fields which can be obtained by putting $e=0$ in Eq.(55) of
\cite{Kruglov2}. As a result, we obtain the matrix equation for
interacting torsion fields
\begin{equation}
\biggl [\alpha _\mu \partial_\mu +\frac 12\left( \sigma _0P_0+\sigma
_1\overline{P} +\sigma _2\overline{\overline{P}}\right) \alpha _{\mu
\nu }\mathcal{F}_{\mu \nu }+m P_1+\frac{m^2_0}{m} P_0\biggr ]\Psi
(x)=0,  \label{32}
\end{equation}
where the projection operators $\overline{P}$,
$\overline{\overline{P} }$, and $\alpha _{\mu \nu }$ being
\begin{equation}
\overline{P}=\varepsilon ^{\mu ,\mu },
\hspace{0.3in}\overline{\overline{P}}=\frac 12\varepsilon ^{[\mu \nu
],[\mu \nu ]},\hspace{0.3in} \alpha _{\mu \nu }=\alpha _\mu \alpha
_\nu -\alpha _\nu \alpha _\mu. \label{33}
\end{equation}
The matrix equation (32) is equivalent to the system of tensor
equations
\[
\partial_\mu S_\mu +\frac{m_0^2}{m^2}S+\frac{\sigma
_0}{m}\mathcal{F}_{\mu\nu}S_{\mu\nu}=0,
\]
\begin{equation}
S_{\mu \nu }-\partial_\mu S_\nu +\partial_\nu
S_\mu+\frac{\sigma_2}{m}\left(\mathcal{F}_{\nu\rho}S_{\mu\rho}-
\mathcal{F}_{\mu\rho}S_{\nu\rho}\right)=0, \label{34}
\end{equation}
\[
\partial_\nu S_{\mu \nu }+\partial_\mu S+m^2 S_\mu
+2\sigma_1m\mathcal{F}_{\mu\nu}S_\nu=0.
\]
The interaction introduced may be considered as an effective
interaction which follows from the fundamental theory. The constants
$\sigma_0$, $\sigma_1$, $\sigma_2$ are connected with the internal
characteristics of torsion particles \cite{Kruglov2}. It means that
if the interaction considered exists, torsion particles may be
treated as composite particles. Possible torsion interaction with
electromagnetic fields was discussed in \cite{Itin}.

\subsection{Hamiltonian form }

Let us consider the Hamiltonian form of Eq.(32) for torsion
particles in the external electromagnetic field. Introducing
dynamical and auxiliary components of the wave function $\alpha_4^2
\Psi (x)=\varphi (x)$, $(1-\alpha_4^2) \Psi (x)=\chi (x)$
\cite{Kruglov2}, we arrive, after exclusion $\chi (x)$, at the
Hamiltonian form:
\[
i\partial_t\varphi(x)=\alpha_4\biggl\{\frac{m_0^2-m^2} {m}P_0
\]
\begin{equation}
+\biggl[\left( 1-\gamma \right)\left(m+\alpha
_a\partial_a\right)+\frac{\sigma_0}{2} P_0\alpha_{\mu \nu
}\mathcal{F}_{\mu \nu }
\end{equation}\label{35}
\[
+\frac{\sigma_1}{2} \overline{P}\alpha _{\mu \nu }\mathcal{F}_{\mu
\nu }+\frac{\sigma_2}{2}\overline{\overline{P}}\alpha _{\mu \nu
}\mathcal{F}_{\mu \nu }\biggr]\left(1 -\frac{1}{m}\Pi
\alpha_a\partial_a\right)\biggr \}\varphi (x),
\]
where $\Pi=1-\alpha_4^2$, $\gamma =(\sigma _2/2m) \Pi
\overline{\overline{P}}\alpha _{\mu \nu }\mathcal{F}_{\mu \nu }$. We
use here the smallness of $\sigma_2$, $\sigma_0$ (see
\cite{Kruglov2}).

The wave function $\varphi (x)$ possesses eight non-zero components
corresponding to states with spin-1 and spin-0 with two values of
the energy. It should be mentioned that states with spin-1 and
spin-0 can not be treated as separate particles, but they are states
of one field with multi-spin 1,0 \cite{Kruglov4}.

\section{Conclusion}

The theory of torsion fields having two spin states, one and zero,
may be treated as the effective theory. The formulation of wave
equations in the first order formalism allows us to separate states
with spin-1 and spin-0 in the covariant manner with help of
projection operators. In the framework of the first order formalism
the Lagrangian and the energy-momentum tensor are found by the
standard procedure. The possible effective interaction of torsion
fields with external electromagnetic fields is discussed and the
Hamiltonian form of an equation is given.

The main difficulty, however, is that the spin-0 state breaks
unitarity. This is because the spin-0 state gives negative
contribution to the Hamiltonian \cite{Kruglov3}. But even without
torsion, the higher derivative quantum gravity suffers this
shortcoming. We can consider the spin-0 state as a ghost and to
remove it, one has to make the limit $m_0 \rightarrow \infty$.


\begin{thebibliography}{999}

\bibitem{Stelle}  K. S. Stelle, Phys. Rev. \textbf{D16}, 953 (1977).

\bibitem{Kruglov} S. I. Kruglov, Ann. Fond. Broglie \textbf{31}, 343,
(2006) (arXiv: hep-th/0606128).

\bibitem{Dvoeglazov} V. V. Dvoeglazov, Int. J. Theor. Phys.
\textbf{37}, 1009 (1998) (arXiv: hep-th/9710159 ); Ann. Fond.
Broglie \textbf{25}, 81 (2000) (arXiv: hep-th/9906083); Hadronic J.
\textbf{26}, 299 (2003) (arXiv: hep-th/0208159).

\bibitem{Kruglov1} S. I. Kruglov, Ann. Fond. Broglie  \textbf{29},
1005 (2004), Erratum-ibid \textbf{31}, 489, (2006)
(arXiv:quant-ph/0408056); Electron. J. Theor. Phys. \textbf{10}, 11,
(2006) (arXiv: hep-ph/0603181); Hadronic J. \textbf{29}, 637, (2006)
(arXiv: hep-ph/0510103); Can. J. Phys. \textbf{85}, 887, (2007)
(arXiv:hep-ph/0507027).

\bibitem{Hehl} F. W. Hehl, P. von der Heyde, and G. D. Kerlick, Rev.
Mod. Phys. \textbf{48}, 393 (1976).

\bibitem{Hehl1} F. W. Hehl, J. D. McCrea, E. W. Mielke, and Y. Ne'eman, Phys.
Rep. \textbf{258}, 1 (1995) (arXiv: gr-qc/9402012).

\bibitem{Capozziello} S. Capozziello, G. Lambiase, and C.
Stornaiolo, Ann. Phys. (Leipzig) \textbf{10}, 713 (2001)
(arXiv:gr-qc/0101038).

\bibitem{Buchbinder} I. L. Buchbinder, S. D. Odintsov, I. L.
Shapiro, Effective action in quantum gravity (IOP Publishing Ltd.,
Btistol, UK, 1992).

\bibitem{Shapiro} I. L. Shapiro, Phys. Rep. \textbf{357}, 113
(2002) (arXiv: hep-th/0103093).

\bibitem{Shapiro1} A. S. Belyaev, I. L. Shapiro, and M. A. B. do Vale,
Phys. Rev. \textbf{D75}, 034014 (2007) (arXiv: hep-ph/0701002).

\bibitem{Carroll} S. M. Carroll, G. B. Field, Phys. Rev.
\textbf{D50}, 3867 (1994) (arXiv: gr-qc/9403058).

\bibitem{Kruglov2} S. I. Kruglov, Int. J. Mod. Phys. \textbf{A16},
4925 (2001) (arXiv: hep-th/0110083).

\bibitem{Kruglov3} S. I. Kruglov, Hadronic J. \textbf{24}, 167
(2001) (arXiv: quant-ph/0110042).

\bibitem{Dvoeglazov1} V. V. Dvoeglazov, Int. J. Mod. Phys. \textbf{B20},
1317 (2006).

\bibitem{Itin} Y. Itin and F. W. Hehl, Phys. Rev.
\textbf{D68}, 127701 (2003) (arXiv: gr-qc/0307063).

\bibitem{Kruglov4} S. I. Kruglov, Symmetry and Electromagnetic Interaction of Fields
with Multi-Spin (Nova Science Publishers, Huntington, New York,
2001).

\end{thebibliography}
\end{document}